\documentclass[aps,prl,twocolumn,showpacs,superscriptaddress]{revtex4}
\usepackage{graphicx}
\begin{document}


\title{Muon spin rotation/relaxation measurements of the non-centrosymmetric superconductor Mg$_{10}$Ir$_{19}$B$_{16}$} 
     \author{A.~A.~Aczel}
     \affiliation{Department of Physics and Astronomy, McMaster University, Hamilton, Ontario, Canada, L8S 4M1}
     \author{T.J.~Williams}
     \affiliation{Department of Physics and Astronomy, McMaster University, Hamilton, Ontario, Canada, L8S 4M1}
     \author{T.~Goko}
     \affiliation{Department of Physics and Astronomy, McMaster University, Hamilton, Ontario, Canada, L8S 4M1}  
     \affiliation{TRIUMF, 4004 Wesbrook Mall, Vancouver, British Columbia, Canada, V6T 2A3}
     \affiliation{Department of Physics, Columbia University, New York, NY 10027, USA}   
     \author{J.P.~Carlo}
     \affiliation{Department of Physics, Columbia University, New York, NY 10027, USA} 
     \author{W.~Yu}
     \affiliation{Department of Physics, Renmin University of China, Beijing 100872, China}
     \author{Y.J.~Uemura}
     \affiliation{Department of Physics, Columbia University, New York, NY 10027, USA} 
     \author{T. Klimczuk}
     \affiliation{Los Alamos National Laboratory, Los Alamos, NM 87545, USA}
     \affiliation{European Commission, Joint Research Center, Institute for Transuranium Elements, Post Box 2340, D-76125 Karlsruhe, Germany}
     \author{J.D. Thompson}
     \affiliation{Los Alamos National Laboratory, Los Alamos, NM 87545, USA}
     \author{R.J.~Cava}
     \affiliation{Department of Chemistry, Princeton University, Princeton, NJ 08544, USA}
     \author{G.~M.~Luke}
     \affiliation{Department of Physics and Astronomy, McMaster University, Hamilton, Ontario, Canada, L8S 4M1}
     \affiliation{Brockhouse Institute for Materials Research, McMaster University, Hamilton, Ontario, Canada, L8S 4M1} 
     \affiliation{Canadian Institute of Advanced Research, Toronto, Ontario, Canada, M5G 1Z8}

\date{\today}

\begin{abstract}
We have searched for time-reversal symmetry breaking fields in the non-centrosymmetric superconductor Mg$_{10}$Ir$_{19}$B$_{16}$ via muon spin relaxation 
in zero applied field, and we measured the temperature dependence of the superfluid density by muon spin rotation in transverse field to investigate the superconducting 
pairing symmetry in two polycrystalline samples of signficantly different purities. In the high purity sample, we detected no time-reversal symmetry breaking fields greater than 
0.05 G. The superfluid density was also found to be exponentially-flat as T$\rightarrow$0, and so can be fit to a single-gap BCS model. In contrast, the lower purity
sample showed an increase in the zero-field $\mu$SR relaxation rate below T$_c$ corresponding to a characteristic field strength of 0.6 G. While the 
temperature-dependence of the superfluid density was also found to be consistent with a single-gap BCS model, the magnitude as  T$\rightarrow$0 was found
to be much lower for a given applied field than in the case of the high purity sample. These findings suggest that the dominant pairing symmetry in 
high quality Mg$_{10}$Ir$_{19}$B$_{16}$ samples corresponds to the spin-singlet channel, while sample quality drastically affects the superconducting 
properties of this system. 
\end{abstract}

\pacs{
76.75.+i, 
74.70.Dd 
}
\maketitle

The spin-singlet Cooper pairs of conventional superconductors require the corresponding superconducting order parameter to exhibit time-reversal 
symmetry\cite{59_anderson}. For spin-triplet pairing, time-reversal symmetry is characteristic of only some of the possible superconducting states, but 
Anderson originally hypothesized that these should all have spatial inversion symmetry\cite{84_anderson}. This then suggested that a material lacking an 
inversion center should be an unlikely candidate for spin-triplet pairing. However, a potential exception was found recently with the discovery of 
superconductivity in the non-centrosymmetric, heavy fermion compound CePt$_3$Si\cite{04_bauer} where the upper critical field was found to exceed the 
paramagnetic limiting field in this material, suggesting possible spin-triplet pairing. A mixed spin-singlet and triplet pairing state was proposed to 
resolve this controversy, and this exotic possibility has stimulated significant effort in both theoretical and experimental studies to understand 
superconductors lacking inversion centers. 

The proposed mixed pairing state was found to be a consequence of superconductivity in a material that exhibits spin-orbit coupling (SOC) but lacks 
inversion symmetry. In this case, the usually doubly-degenerate electronic bands become non-degenerate almost everywhere in the Brillouin zone and the 
resulting superconducting states can no longer be classified according to parity. While non-centrosymmetric, heavy fermion 
superconductors\cite{04_bauer, 04_akazawa, 05_kimura, 04_sugitani} have been studied in more detail to-date, these materials often have 
magnetically-ordered states that compete or coexist with the superconducting state, and such behaviour tends to significantly complicate penetration depth
measurements. For example, the heavy fermion superconductor CePt$_3$Si first becomes magnetically-ordered at 2.2~K before superconducting below 
0.75~K\cite{04_bauer}, and the two states have been shown to coexist on a microscopic level\cite{04_metoki}. Transition-metal compounds 
such as Li$_2$M$_3$B (M~=~Pd, and Pt)\cite{04_togano, 05_badica}, Mg$_{10}$Ir$_{19}$B$_{16}$\cite{06_klimczuk}, and M$_2$Ga$_9$ 
(M~=~Rh, Ir)\cite{07_shibayama} are therefore generally more straightforward systems for exploring the effect of SOC on superconductivity in the absence of
inversion symmetry.  

The strength of the SOC in any material is determined by both the crystallographic structure and the elemental 
composition. The latter is well-illustrated through recent NMR\cite{07_nishiyama} and magnetic penetration depth\cite{06_yuan} measurements on Li$_2$M$_3$B, 
which found that increasing SOC by replacing Pd with Pt changes the superconducting order parameter from dominantly spin-singlet (Li$_2$Pd$_3$B) to nodal, 
spin-triplet (Li$_2$Pt$_3$B). Muon spin rotation measurements of the penetration depth in Li$_2$Pd$_3$B support the conclusion that Li$_2$Pd$_3$B is 
predominantly an s-wave BCS superconductor with only one isotropic energy gap\cite{06_khasanov}. 

Mg$_{10}$Ir$_{19}$B$_{16}$ crystallizes in a non-centrosymmetric, body-centered cubic structure with space group I$\bar{4}$3m and lattice constant 
a$=$10.568 \AA\cite{09_xu}.  A combination of resistivity, specific heat, and magnetization measurements indicate that this material is a superconductor 
with T$_c$ ranging between 4.5 and 5~K depending on the exact sample investigated\cite{06_klimczuk, 07_klimczuk}. Since Ir is a heavy transition metal SOC is expected to be strong in 
this material, and so questions about the superconducting pairing symmetry are of significant interest. 

The superconducting pairing symmetry of Mg$_{10}$Ir$_{19}$B$_{16}$ has already been investigated through both heat capacity\cite{07_klimczuk, 07_mu} and photoemission 
spectroscopy\cite{09_yoshida, 09_yoshida_2}, and the data is consistent with isotropic s-wave pairing in both cases. However, point contact 
spectroscopy\cite{07_klimczuk} and penetration depth measurements performed using a tunnel diode oscillator\cite{09_bonalde} suggest that the system is 
better described by two-gap superconductivity. The two-gap model is consistent with spin-split bands that are a direct consequence of the lack of inversion
symmetry combined with significant SOC, and is expected in the case of a mixed pairing state with a significant spin-triplet component. To help resolve 
the apparent disagreement regarding the nature of the superconducting pairing symmetry in Mg$_{10}$Ir$_{19}$B$_{16}$, more experiments need to be 
performed. 

Muon spin rotation is a powerful technique for measuring the penetration depth and 
determining the nature of the pairing symmetry in superconductors. Due to maximal parity violation in the weak interaction decay of pions, the muons 
produced for such an experiment are initially $\sim$100\% polarized. Muon spin rotation (TF-$\mu$SR) employs a transverse-field geometry, as the muons are 
implanted into the sample one at a time with the initial polarization oriented perpendicular to the applied magnetic field. The muon spins exhibit Larmor 
precession about the local magnetic field and then decay into a positron, plus two neutrinos which are not detected. The resulting positron travels 
preferentially in the direction of the muon spin at the time of decay, and so due to the asymmetric nature of this decay one can track the polarization of 
the muon ensemble as a function of time by setting up positron counters around the sample. In type II superconductors, the mixed (or vortex) state gives 
rise to a spatial distribution of local magnetic fields; this field distribution manifests itself in the $\mu$SR signal through a relaxation of the muon 
polarization. $\mu$SR has been used successfully for some time to study the mixed state in superconductors; further details can be found in 
\cite{07_sonier, 07_russo}. 

Zero field (ZF) $\mu$SR is very complementary to TF-$\mu$SR superconductor studies. The experimental set-up is essentially the same, but no field is 
applied. $\mu$SR is an extremely sensitive probe of magnetism (internal fields on the order of 0.1~G can be readily detected), and so when investigating 
superconductors a ZF-$\mu$SR experiment is generally performed in advance of TF-$\mu$SR measurements to search for any anomalous magnetism that was not 
previously detected by other techniques. This knowledge is quite important since the presence of magnetism in a superconductor can significantly 
complicate penetration depth measurements. 

\begin{figure}[t]
\includegraphics[width=2.7in,angle=0]{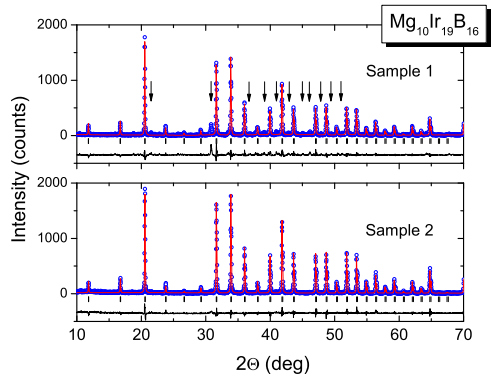}
\caption{\label{fig1} Powder x-ray diffraction measurements of the two Mg$_{10}$Ir$_{19}$B$_{16}$ samples used for the $\mu$SR measurements. The arrows indicate 
the dominant impurity peaks in the sample 1 pattern.}
\end{figure}

ZF-$\mu$SR is also useful to search for time-reversal symmetry-breaking fields; the presence of such fields limits the possible superconducting states and 
the associated pairing symmetry. For example, time-reversal symmetry is a prerequisite for any spin-singlet state, but this symmetry is broken for certain 
spin-triplet states, specifically for those which have a degenerate representation. The presence of two or more degenerate superconducting phases naturally 
leads to a spatially-inhomogeneous order parameter near the resulting domain walls; this creates spontaneous supercurrents and hence magnetic fields near 
those regions. Another possible origin of time-reversal symmetry-breaking fields is from intrinsic magnetic moments due to spin polarization (for spin 
triplet pairing) and the relative angular momentum of the Cooper pairs\cite{91_sigrist}. Broken time-reversal symmetry was previously detected in
the unconventional superconductor Sr$_2$RuO$_4$ via $\mu$SR\cite{98_luke}, as a local field of $\sim$~0.5~G was found below T$_c$ at the muon site. This 
provided some of the strongest evidence that Sr$_2$RuO$_4$ is actually a p-wave superconductor. Similar time-reversal symmetry-breaking fields could be 
detected by $\mu$SR if the spin-triplet contribution to the superconducting state in Mg$_{10}$Ir$_{19}$B$_{16}$ is significant.
     
In this work, we have used both ZF and TF-$\mu$SR to investigate the nature of the superconductivity in two polycrystalline Mg$_{10}$Ir$_{19}$B$_{16}$ samples of different quality. The former allowed 
us to search for time-reversal symmetry-breaking fields, and the latter allowed us to measure the penetration depth and characterize the pairing symmetry. 
Our data from the high-quality sample is consistent with a single-gap s-wave model; this is in agreement with the earlier heat capacity\cite{07_mu} and photoemission spectroscopy 
work\cite{09_yoshida, 09_yoshida_2}.

Mg$_{10}$Ir$_{19}$B$_{16}$ samples were synthesized by a standard solid state reaction of pure Mg, Ir, and B elements as described 
elsewhere, and the chemical composition was accurately determined via electron diffraction\cite{06_klimczuk}. Prior to the $\mu$SR measurements, the samples were characterized by powder 
x-ray diffraction, performed on a Scintag XDS 2000 diffractometer with Cu-K$_\alpha$ radiation ($\lambda$~$=$0.15460 nm). Rietveld refinements of the structures were achieved using 
GSAS\cite{01_toby, 00_larson}, and slightly different lattice constants of 10.5629(4)~\AA~and 10.5655(1)~\AA~were found for sample 1 and 2 respectively. This is likely due to more vacancies in the 
crystal structure of the former, and when one also considers that many more impurity peaks are observed in the corresponding x-ray pattern (shown in Fig. 1), this suggests that sample 2 is a much higher 
quality specimen with fewer impurities/defects. The magnetic susceptibility of each sample was also measured with a SQUID magnetometer; the zero-field cooled (ZFC) and 
field-cooled (FC) measurements, shown in Fig. 2, confirmed nice superconducting properties. If we let T$_c$ correspond to the onset of the susceptibility drop, then  
T$_c$~$\sim$~4.8 K for sample 1 and 4.6 K for sample 2. The much sharper drop below T$_c$, observed in both the FC and ZFC measurements of sample 2, provides further evidence for the better 
sample quality in this case. 

\begin{figure}[t]
\includegraphics[width=3.4in,angle=0]{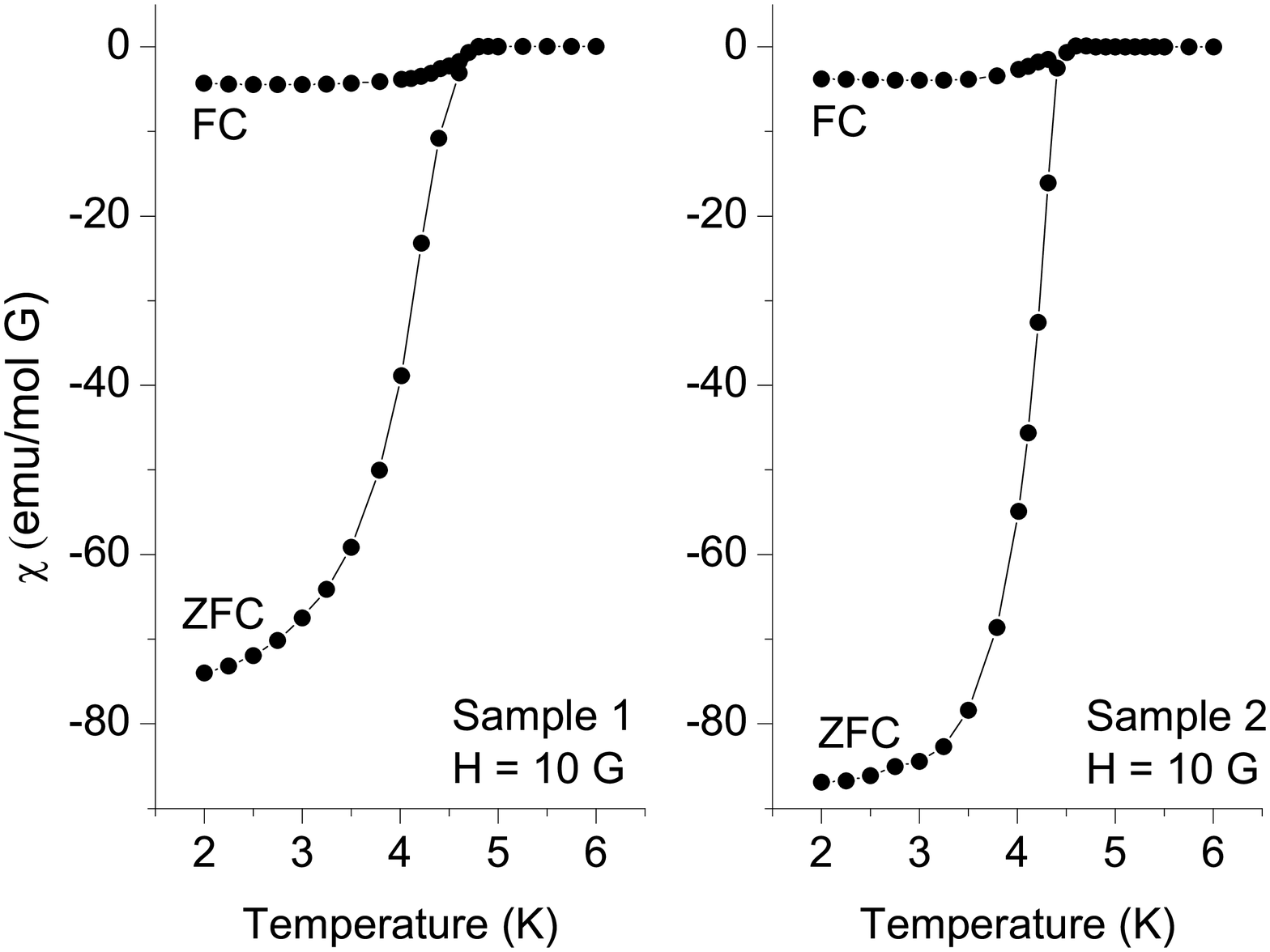}
\caption{\label{fig2}ZFC and FC susceptibility measurements of the two Mg$_{10}$Ir$_{19}$B$_{16}$ samples used for the $\mu$SR measurements.}
\end{figure}

\begin{figure}[t]
\includegraphics[width=3.4in,angle=0]{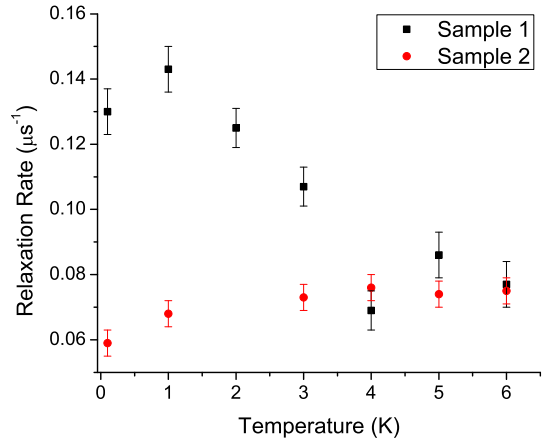}
\caption{\label{fig3}ZF-$\mu$SR relaxation rate vs. temperature. The relaxation rate of sample 1 shows a small increase around T$_c$, while
the relaxation rate of sample 2 remains essentially constant within the error bars. Possible reasons for this discrepancy are outlined in the text.}
\end{figure}

ZF-$\mu$SR data was collected using the dilution refrigerator in conjunction with the M15 surface muon 
beamline at TRIUMF. Fig. \ref{fig3} depicts the ZF muon spin exponential relaxation rate vs. temperature for our two Mg$_{10}$Ir$_{19}$B$_{16}$ samples. 
Note that the data was fit to the function 
\begin{equation}
A(t)=A_0e^{-\lambda t}G_{KT}(t)
\end{equation}
where G$_{KT}$(t) is the Gaussian Kubo-Toyabe relaxation function given by:
\begin{equation}
G_{KT}(t)=\frac{1}{3}+\frac{2}{3}(1-\Delta^2t^2)e^{-\frac{1}{2}\Delta^2t^2}
\end{equation}
The Gaussian Kubo-Toyabe relaxation is due to nuclear dipolar fields and is temperature-independent; we determined this to be $\Delta$~$=$~0.207(4)~$\mu$s$^{-1}$ from a global 
fit incorporating all of our ZF data for the two samples. The exponential relaxation rate was set as a variable parameter, and this is plotted as a function of temperature for the two samples in Fig. 3. 
Sample 1 shows a small increase in the exponential relaxation rate of $\sim$0.05~$\mu$s below T$_c$, corresponding to a characteristic field strength of B~$\sim$~0.6~G. However, the exponential 
relaxation rate of sample 2 remains essentially constant below T$_c$. Although we cannot completely rule out the presence of time-reversal symmetry breaking fields in sample 2, the 
current data does set an upper bound on the magnitude of these fields to be $\sim$~0.05~G. 

There are a few different scenarios that can explain the discrepancy between the ZF data of the two samples. One possibility is that the relaxation 
increase in sample 1 is actually due to time-reversal symmetry-breaking fields. The presence of such fields suggests that the associated superconducting
state is at least two-fold degenerate, and the spatially-varying order parameter would then lead to the presence of supercurrents in the bulk of the superconductor\cite{91_sigrist}.
If this is the case, one would generally expect these fields to exist in sample 2 also, although we note that the domain size may be substantially smaller in sample 1 due to the poorer quality. Many more muons would then stop near the 
domain walls in sample 1 and see a non-zero local field associated with the currents flowing at the domain boundaries; this could lead to the increased broadening below T$_c$ for this case only. Furthermore, NMR and electron diffraction
measurements were performed very recently on two similar samples with compositions Mg$_{9.3}$Ir$_{19}$B$_{16.7}$ and Mg$_{10.5}$Ir$_{19}$B$_{17.1}$\cite{09_tahara}. The electron diffraction measurements clearly confirmed that 
the former was a better quality sample with fewer defects/impurities, and the NMR measurements indicated that only the other sample had a finite spin susceptibility in the zero temperature limit. These results
then suggested that defects/impurities may enhance spin orbit coupling in this system thus leading to an increased spin-triplet component of the superconductivity; this interpretation is also consistent with the observation of time-reversal 
symmetry-breaking fields in our lower quality sample only. 

There are also other possible explanations for the difference between the ZF data of the two samples not related to time-reversal symmetry-breaking fields. For example, the increased ZF relaxation in sample 1
could be due to dilute magnetic impurities throughout the sample. However, this seems rather unlikely since the onset of the increased relaxation rate 
occurs very close to T$_c$ and furthermore, these are not expected in this system. The other possibility is related to field-zeroing. If the applied field was not perfectly zeroed in our experiment, there 
could potentially be some trapped flux in the superconducting state of these samples. Sample 1 may be more efficient at trapping flux than sample 2 due to the lower quality of the former, and 
this could result in an increased relaxation below T$_c$ for sample 1 only. We note that the ZF measurements were conducted on both samples back-to-back without ever 
applying a field during this time, and so any residual applied field should be the same for both samples. 

While it is not possible for us to distinguish between these different possibilities, we can conclude with certainty that a high purity sample of Mg$_{10}$Ir$_{19}$B$_{16}$ showed no evidence for time reversal symmetry-breaking fields greater than 0.05~G. We also observed only very weak ZF relaxation in both samples at all temperatures, and so this allows for reliable penetration depth measurements to be performed.  

\begin{figure}[t]
\includegraphics[width=3.5in,angle=0]{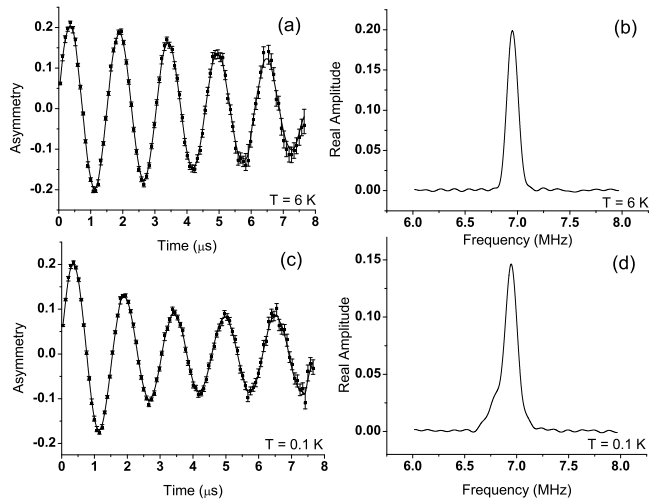}
\caption{\label{fig4}Mg$_{10}$Ir$_{19}$B$_{16}$ sample 1 asymmetry vs. time and the corresponding fast-Fourier transforms at TF~$=$~500~G 
for both 6 K [shown in (a) and (b)] and 0.1 K [shown in (c) and (d)]. While the increased broadening is apparent for T~$=$~0.1~K, the characteristic 
Abrikosov lineshape was not observed. This is likely due to the large background signal superimposed on the data.}
\end{figure}

With this in mind we proceeded to collect TF-$\mu$SR data, as measuring the penetration depth would allow us to learn more about the nature of the 
superconducting pairing symmetry. Note that all of the data was collected via field cooling. Fig. \ref{fig4} shows the Fast-Fourier transforms (FFTs) 
obtained at both 6 and 0.1~K. Below T$_c$, the characteristic lineshape of an Abrikosov vortex lattice was not resolved as a large background signal from 
the silver sample holder was superimposed on the data. However, we did find significant line-broadening below T$_c$ as expected for a superconductor (see 
Fig. \ref{fig4}). This broadening was analyzed as a Gaussian field-distribution; the width of this distribution $\sigma_{sc}$ has been shown 
previously to be proportional to the superfluid density $\sigma_{sc}~\propto~n_s/m^\ast~\propto~1/\lambda^2$\cite{88_brandt}. $\sigma_{sc}$ was first 
obtained by fitting the $\mu$SR data in the time domain to the functional form:
\begin{equation}
A(t)=A_0e^{-\sigma^2t^2/2}cos(\omega t+\phi)+A_{bg}cos(\omega_{bg} t+\phi)
\end{equation}
The non-relaxing background term was included to account for the muons stopping in the silver sample holder. The first term gives the total 
sample relaxation rate $\sigma$; there are contributions from both the vortex lattice and nuclear dipole moments below T$_c$. 
The contribution from the vortex lattice, $\sigma_{sc}$, was determined by quadratically-subtracting the background nuclear dipolar relaxation rate 
obtained from spectra measured above T$_c$. 

\begin{figure}[t]
\includegraphics[width=3.7in,angle=0]{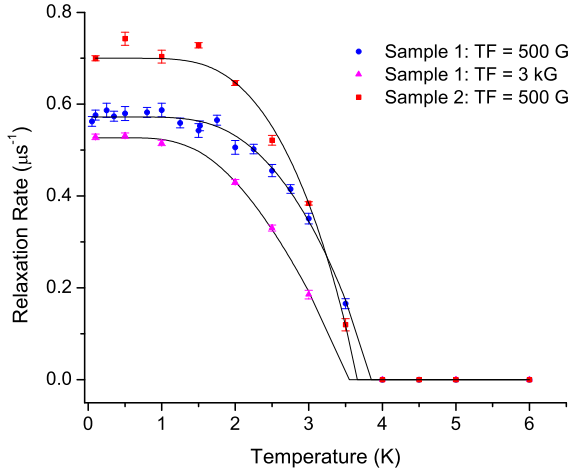}
\caption{\label{fig5}TF-$\mu$SR superconducting relaxation rate $\sigma_{sc}$ vs. temperature. The data can be adequately fit to a single-gap BCS model, 
as indicated by the solid lines.}
\end{figure}

Fig. \ref{fig5} depicts the TF muon spin relaxation rate (proportional to superfluid density) vs. temperature for our two Mg$_{10}$Ir$_{19}$B$_{16}$ 
samples. Sample 1 was investigated with both TF~$=$~500~G and 3~kG, while sample 2 was investigated only in 500~G. The sample 1 data indicates that the 
superfluid density decreases with increasing applied field as expected. The superfluid density of the two samples at 500 G as T$\rightarrow$ 0 is 
noticeably different; sample 2 has a higher superfluid density than sample 1 and this is consistent with the higher quality of the former. However, the temperature dependence 
of the superfluid density is the same for both samples: namely, the superfluid density seems to saturate as T~$\rightarrow$~0. Since conventional BCS theory predicts an 
exponentially-flat temperature-dependence of the superfluid density at low temperatures, we proceeded to fit our data to this model. The agreement is quite good for both 
samples as seen in Fig. \ref{fig5}, and so this suggests that the pairing symmetry in this material can likely be described by a single-gap s-wave model. 
For TF~$=$~500~G, fitting to this model gave us the following parameters: 2$\Delta$/kT$_c$~$=$~4.74(9) (2$\Delta$~$=$~1.6~meV), T$_c$~$=$~3.8(1)~K, and 
$\lambda$(T$\rightarrow$ 0)~$=$~3570(20)~\AA~for sample 1 and 2$\Delta$/kT$_c$~$=$~4.93(5) (2$\Delta$~$=$~1.6~meV), T$_c$~$=$~3.7(1)~K and 
$\lambda$(T$\rightarrow$ 0)~$=$~3230(20)~\AA~for sample 2. Note that our 2$\Delta$/kT$_c$ values are only slightly larger than what one expects for 
weak-coupled BCS theory, and that the penetration depth values were obtained from $\sigma_{sc}$ through a conversion procedure outlined in \cite{91_uemura}. The 
significantly suppressed T$_c$'s in only a modest applied field of 500~G are consistent with the relatively low H$_{c2}\sim$8~kG\cite{07_klimczuk} of this system. 

Motivated by the earlier work proposing multigap superconductivity in Mg$_{10}$Ir$_{19}$B$_{16}$\cite{07_klimczuk, 09_bonalde}, we also tried to fit our 
data to a phenomenological two-gap model as described elsewhere\cite{09_williams}. Since the fitting routine assigned a weighting of 0 (within our error 
bars) to one of the gaps, we found that a two-gap model does not seem to describe our data unless the two gaps are very similar in magnitude. However, we 
note that previous measurements of Mg$_{10}$Ir$_{19}$B$_{16}$ suggesting multigap superconductivity were essentially performed in ZF, while our $\mu$SR measurements 
were performed in an applied field. While a recent $\mu$SR penetration depth measurement performed by our group on the superconductor 
Ba(Fe$_{0.926}$Co$_{0.074}$)$_2$As$_2$ found evidence for multigap superconductivity in that material, the same measurements also revealed that the relative weighting 
factor of the smaller gap quickly approaches 0 with increasing TF\cite{09_williams}. This could also be applicable to the present case. 

Although the above results come with the warning 
that previous $\mu$SR measurements on polycrystalline cuprate samples have resulted in the correct temperature-dependence of the superfluid density being 
masked, this may be less of a concern in the present case due to the cubic crystal symmetry and hence the isotropic penetration depth. In any event, future 
penetration depth measurements should be conducted on single crystals when they become available to confirm our results.

In conclusion, we have used both ZF and TF-$\mu$SR to investigate the superconductivity in two samples of the non-centrosymmetric system Mg$_{10}$Ir$_{19}$B$_{16}$. 
In the higher quality sample, ZF-$\mu$SR measurements find no evidence for time-reversal symmetry-breaking fields larger than 0.05~G. TF-$\mu$SR measurements of both
samples yield a superfluid density that is exponentially-flat at low temperatures, and so our data can be fit to a single-gap BCS model. These findings suggest that high quality Mg$_{10}$Ir$_{19}$B$_{16}$
exhibits spin-singlet s-wave pairing, even though the lack of inversion symmetry and the large expected SOC of Mg$_{10}$Ir$_{19}$B$_{16}$ should yield a mixed pairing state, presumably with a large 
spin-triplet contribution. Our work also suggests that the superconducting properties of this system are very sensitive to sample quality, and so further studies are required to investigate the roles of both 
SOC and defects/impurities on the superconductivity of this material in more detail.

\begin{acknowledgments}
Research at McMaster University is supported by NSERC and CIFAR. Research at Columbia University is supported by the NSF under contracts No. 
DMR-05-02706 and No. DMR-08-06846 (Material World Network). Work at Los Alamos was performed under the auspices of the US Department of Energy and the Office of Basic Energy Sciences.
\end{acknowledgments}

\vfill \eject
\end{document}